\documentclass[journal]{IEEEtran}

\usepackage{amsmath}
\usepackage{epsfig}
\usepackage{amssymb}
\usepackage{graphicx}
\usepackage{subfigure}
\usepackage{stfloats}

\usepackage{multirow}

\begin{document}

\title{Comprehensive Analysis and Measurement of Frequency-Tuned and Impedance-Tuned Wireless Non-Radiative Power Transfer Systems}

\author{Jason~D.~Heebl,~\IEEEmembership{Member,~IEEE,}
        Erin~M.~Thomas,~\IEEEmembership{Member,~IEEE,}
        Robert P. Penno,~\IEEEmembership{Senior Member,~IEEE,}
        and Anthony Grbic,~\IEEEmembership{Member,~IEEE}

\thanks{J. D. Heebl is with the Space and Missiles Systems Center, Satellite Control Network, Los Angeles Air Force Base, El Segundo, CA 90245 USA (email: Jason.Heebl.1@us.af.mil).}
\thanks{E. M. Thomas is with SRI International in the Communications, Radar and Sensing Lab.}
\thanks{R. P. Penno is with the Department of Electrical and Computer Engineering, University of Dayton, Dayton, OH 45469 USA.}
\thanks{A. Grbic is with the Radiation Laboratory, Department of Electrical Engineering and Computer Science, University of Michigan, Ann Arbor, MI 48109-2122 USA (email: agrbic@umich.edu).}
}

\maketitle

\begin{abstract}
This paper theoretically and experimentally investigates frequency-tuned and impedance-tuned wireless non-radiative power transfer (WNPT) systems. Closed-form expressions for the efficiencies of both systems, as a function of frequency and system (circuit) parameters, are presented. In the frequency-tuned system, the operating frequency is adjusted to compensate for changes in mutual inductance that occur for variations of transmitter and receiver loop positions.  Frequency-tuning is employed for a range of distances over which the loops are strongly coupled.  In contrast, the impedance-tuned system employs varactor-based matching networks to compensate for changes in mutual inductance and achieve a simultaneous conjugate impedance match over a range of distances. The frequency-tuned system is simpler to implement, while the impedance-tuned system is more complex but can achieve higher efficiencies.  Both of the experimental WNPT systems studied employ resonant shielded loops as transmitting and receiving devices.
\end{abstract}

\section{Introduction}

\newcounter{tempequationcounter}
\setcounter{equation}{5}

\begin{figure*}[b!]
\normalsize
\hrulefill
\begin{equation}
\left|\Gamma_{\mathrm{in}}\right|^2 =  \frac{\left[\left(R-R_L\right) \left(R+R_L\right) - \left(\omega L - \frac{1}{\omega C}\right)^2 + \left(\omega M_{12}\right)^2 \right]^2 + 4 R^2 \left(\omega L - \frac{1}{\omega C}\right)^2}
 {\left[\left(R+R_L\right)^2 - \left(\omega L - \frac{1}{\omega C}\right)^2 + \left(\omega M_{12}\right)^2 \right]^2 + 4 \left(R+R_L\right)^2 \left(\omega L - \frac{1}{\omega C}\right)^2}
\label{eq:ReflPower} \\
\end{equation}
\begin{equation}
\eta = \left(1-\left|\Gamma_{in}\right|^2\right) \eta' = \frac{\left[2 R_L \left(\omega M_{12}\right)\right]^2}{\left[\left(R+R_L\right)^2 - \left(\omega L - \frac{1}{\omega C}\right)^2 + \left(\omega M_{12}\right)^2\right]^2 + 4\left(R+R_L\right)^2\left(\omega L - \frac{1}{\omega C}\right)^2}
\label{eq:EffwRefl}
\end{equation}
\vspace*{4pt}
\end{figure*}

\setcounter{equation}{\value{tempequationcounter}}

\IEEEPARstart{A}{ttempts} to transfer power wirelessly have generally relied on transmitting and receiving radiated power \cite{PowerHistory, LowPowerFarField}. In addition to these far-field systems, wireless non-radiative power transfer (WNPT) via quasi-static magnetic fields has garnered significant interest in recent years. Wireless power transfer using conventional magnetic induction as well as resonant magnetic induction for increased range (mid-range distances) are actively being pursued in academia and industry \cite{eCoupled, WPC, A4WP, PMA, InductivePowerTransfer, Coil4ConsumerElec}. WNPT by resonant magnetic induction can trace its roots to the work of Nikola Tesla in the early twentieth century \cite{Tesla-Patent1,Tesla-Patent2,SecorArticle}. Much of its recent popularity, however, can be attributed to work at MIT in 2007, where power was transferred between magnetically-coupled resonant coils \cite{Soljacic}.  The transmitting and receiving coils formed a transformer, and resonance was used to improve power transfer efficiency.  The transformer in such a scheme is non-ideal given that the loop inductances are finite, the coefficient of coupling is notably less than unity, and the loops are lossy \cite{Everitt}.

In \cite{Thomas, Thomas2}, WNPT utilizing two coupled shielded-loop resonators, instead of resonant coils, was reported. Closed-form expressions for the loops' circuit parameters, input and output impedances, and efficiency were developed from a circuit model of the system. Fixed matching networks were used to optimize efficiency for specific distances, that is particular values of mutual inductance. Experimental results showed close agreement with theoretical predictions of optimal efficiency.

In recent years, there have been attempts to develop improved WNPT systems with increased efficiency for a range of distances, or equivalently mutual coupling. Techniques such as tuning the frequency of operation, employing impedance-matching networks \cite{MatchingNet1, MatchingNet2, Multireceive}, transponder configurations \cite{Mauro, CritReview}, coil arrays \cite{PhasedArray}, and the use of a superlens to enhance coupling \cite{Superlens} have been explored. In \cite{Intel,Korea,VehicalPower}, constant efficiencies were maintained by tuning a WNPT system's operating frequency while the transmit and receive resonators were strongly coupled. In \cite{FreqTune-Exp}, the performance of such a frequency-tuned system was compared in simulation with an impedance-matched system. However, analytical models explaining frequency-tuning with fixed source and load impedances have not been developed to date. In addition, experiments that satisfactorily compare the full operation of both approaches have not been reported.

In this paper, we present an analytical investigation of both frequency-tuned and impedance-matched WNPT systems. The governing equations of a WNPT system are investigated under strong, critical, and weak coupling between transmit and receive resonators. Conditions for optimal frequency-tuned operation, with fixed source and load impedances, are presented. The expressions analytically demonstrate why constant efficiencies can be attained in a strongly-coupled system. In addition, it is analytically shown that impedance-matched systems are capable of providing optimal power transfer for all mutual inductance values.

Finally, the analytical findings are experimentally validated for a frequency-tuned and impedance-tuned WNPT system. Frequency-tuning is demonstrated for three fixed matching networks. Each matching network provides a simultaneous conjugate match to source and load impedances at a fixed distance. The system is frequency-tuned for spacings shorter than this fixed distance. In addition, varactor-based matching networks are used to demonstrate an impedance-tuned WNPT system. It is shown that such a system exhibits maximum efficiency. The maximum efficiencies achieved using the impedance-tuned WNPT system  are compared to those of the frequency-tuned system.

\section{Circuit Model}
\label{sec:CircuitModel}

\begin{figure}[htbp]
    \centering
    \includegraphics[width=3.5in]{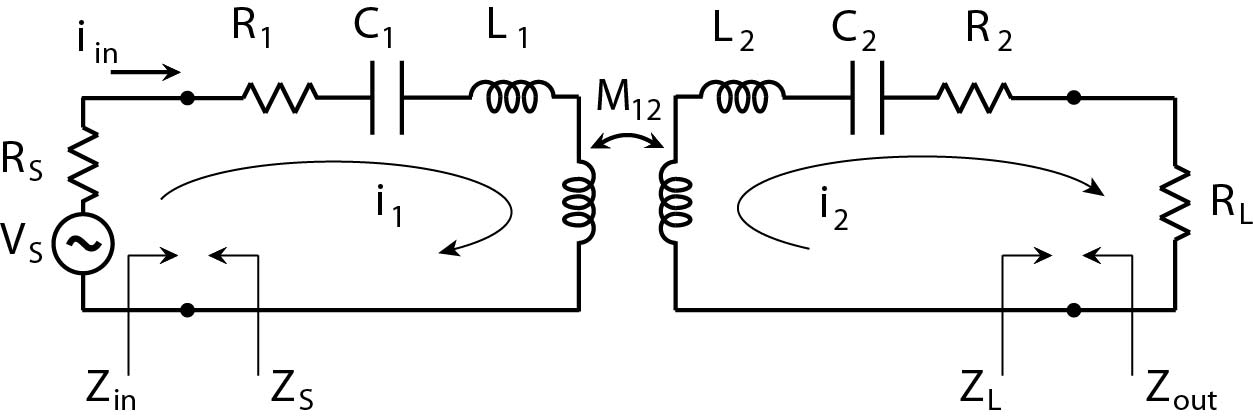}
    \caption{Equivalent circuit for two  inductively coupled resonant loops employed in a wireless non-radiative power transfer system.}
    \label{fig:BasicCkt}
\end{figure}

The electrically-small loops of WNPT systems are typically modeled using series resonator circuits \cite{FundamentalWNPT}. Fig. \ref{fig:BasicCkt} shows the lumped-element circuit model of a WNPT system based on resonant magnetic coupling between two electrically small loops \cite{Thomas, Thomas2}. The variable $M_{12}=M_{21}$ denotes the mutual inductance between two loops.

From this circuit model, the input and output impedances can be found:
\begin{align}
Z_{in}&=j\omega L_1-\frac{j}{\omega C_1} + R_1+\frac{\left(\omega M_{12}\right)^2}{R_2+Z_L+j\omega L_2-\frac{j}{\omega C_2}}
\label{eq:Zin} \\
Z_{out}&=j\omega L_2-\frac{j}{\omega C_2}+R_2+\frac{\left(\omega M_{12}\right)^2}{R_1+Z_S+j\omega L_1-\frac{j}{\omega C_1}}.
\label{eq:Zout}
\end{align}
For simplicity, it will be assumed that the loops depicted in Fig. \ref{fig:BasicCkt} are identical:
\begin{equation}
R_1=R_2=R, L_1=L_2=L, C_1=C_2=C.
\label{eq: identicl}
\end{equation}
The efficiency, $\eta$, will be defined as the ratio of power delivered to the load ($P_L$) to power available from the source ($P_{AVS}$) as follows:
\begin{align}
\eta &= \frac{P_{L}}{P_{AVS}} = \left(1-\left|\Gamma_{\mathrm{in}}\right|^2\right) \eta' \label{eq:Efficiency2a} \\
 &= \left(1-\left|\Gamma_{\mathrm{in}}\right|^2\right) \times \notag \\
 &\frac{ \left(\omega M_{12}\right)^2 R_L}{ R \left[\left(R + R_L\right)^2 + \left(\omega L - \frac{1}{\omega C}\right)^2\right] + \left(\omega M_{12}\right)^2 (R + R_L)}, \label{eq:Efficiency2}
\end{align} where $\Gamma_{\mathrm{in}}=\frac{Z_{in}-Z_S}{Z_{in}+Z_S}$ is the input reflection coefficient. The power available from the source is the maximum power that can be delivered to the network: $P_{AVS} = \frac{|V_S|^2}{8 \Re e\{{Z_{S}}\}}$. It does not depend on the input impedance, $Z_{in}$, of the network \cite{Gonz}. The reflected power at the input is proportional to $\left|\Gamma_{\mathrm{in}}\right|^2$ given by (\ref{eq:ReflPower}).

\setcounter{equation}{7}

When (\ref{eq:ReflPower}) is substituted into (\ref{eq:Efficiency2}), the efficiency equation simplifies to the expression shown in (\ref{eq:EffwRefl}). The efficiency, given by (\ref{eq:Efficiency2}),(\ref{eq:EffwRefl}), is the transducer gain of the two-port network when the source and load impedances are identical, real quantities \cite{Pozar}. The quantity $\eta'$ represents the power gain.

At the resonant frequency of the isolated  loops ($\omega_0 = 1/\sqrt{LC}$), the efficiency (\ref{eq:Efficiency2}) reduces to:
\begin{align}
\eta &= \left(1-\left|\Gamma_{\mathrm{in}}\right|^2\right) \eta' \\
&=  \left(1-\left|\Gamma_{\mathrm{in}}\right|^2\right) \times \notag \\
 &\qquad\qquad\frac{ \left(\omega_0 M_{12}\right)^2 R_L}{(R+R_L) \left[ R\left(R + R_L\right)  + \left(\omega_0 M_{12}\right)^2 \right]}.
\label{eq:Efficiency_Base}
\end{align}  In addition, at $\omega_0$, the square of the reflection coefficient magnitude reduces from (\ref{eq:ReflPower}) to:
\begin{equation}
\left|\Gamma_{\mathrm{in}}\right|^2 = \left[ \frac{R^2-{R_L}^2 + \left(\omega_0 M_{12}\right)^2 }
 {\left(R+R_L\right)^2 + \left(\omega_0 M_{12}\right)^2} \right]^2.
\label{eq:ReflPower_wkCrt}
\end{equation}
Therefore, the complete expression for efficiency at resonance $\omega_0$ simplifies to:
\begin{equation}
\eta = \left(1-\left|\Gamma_{\mathrm{in}}\right|^2\right) \eta' = \left[ \frac{2R_L \left(\omega_0 M_{12}\right)}{\left(R+R_L\right)^2 + \left(\omega_0 M_{12}\right)^2} \right]^2.
\label{eq:Efficiency_WkCrt}
\end{equation}
These expressions account for both dissipation and reflections.

All the equations reported in this section were derived from from the circuit shown in Fig. \ref{fig:BasicCkt}, under the conditions specified. They are in agreement recent papers on the topic \cite{Coil4ConsumerElec, WPTnear2far, Midfield4Implantable, MidrangeWNPTeqs}. The following sections analytically discuss tuning techniques to improve efficiency, and then validate the expressions through experimentation.

\section{Analysis of Tuning Techniques}

Coupled-mode theory asserts that there are three regions of operation for two coupled resonators \cite{Soljacic, Everitt}. These regions are based on the strength of the mutual coupling between the two loop resonators relative to losses:
\begin{itemize}
\item strong coupling: $\left(\omega M_{12}\right)^2 > \left(R+R_L\right)^2$
\item critical coupling: $\left(\omega M_{12}\right)^2 = \left(R+R_L\right)^2$
\item weak coupling: $\left(\omega M_{12}\right)^2 < \left(R+R_L\right)^2$
\end{itemize}
To obtain the resonances of the WNPT system, the imaginary part of the input impedance (\ref{eq:Zin}) is set to zero:
\begin{multline}
0=\left(\omega^2LC - 1\right) \times \\
\left[\left(R+R_L\right)^2 + \left(\omega L -\frac{1}{\omega C}\right)^2 - \left(\omega M_{12}\right)^2\right].
\label{eq:Zin_Res_Cond}
\end{multline}
Equation (\ref{eq:Zin_Res_Cond}) identifies three resonances of the coupled loops: conditions under which $\mathrm{imag}(Z_{in})=0$. The first term defines the resonant frequency, $\omega_0 = \frac{1}{\sqrt{LC}}$, of the loops in isolation. At this frequency, $\omega_0$, the currents in the two loops are 90 degrees out of phase. The expression in square parentheses defines the even and odd mode resonances of the coupled loop system:
\begin{equation}
\left(\omega L -\frac{1}{\omega C}\right)^2 =\left(\omega M_{12}\right)^2 - \left(R+R_L\right)^2.
\label{eq:SplitModes}
\end{equation}
The currents in the two loops are in phase for the even mode, while 180 degrees out of phase for the odd mode.  The even and odd mode resonances separate in frequency as the coupling between the two loops becomes stronger. An approximate expression for the frequency separation between these two modes can be found by setting $\omega = \omega_0 \pm \frac{\Delta \omega}{2}$ and solving (\ref{eq:SplitModes}) for $\Delta \omega$:
\begin{equation}
\Delta \omega \approx \sqrt{\left(\frac{{\omega}_o M_{12}}{L}\right)^2-\left(\frac{R+R_L}{L}\right)^2}.
\label{eq:Freq_Split}
\end{equation}
When the system is strongly coupled, $\Delta \omega$ is positive and real, producing two resonant frequencies: an even and odd mode. At critical coupling, $\Delta \omega =0$ and the even and odd mode frequencies merge to $\omega_0$. When the system is weakly coupled, $\omega =\omega_0 \pm \frac{\Delta \omega}{2}=\omega_0 \pm j\left(\frac{R+R_L}{2L}\right)$. As a result, the resonant frequency remains $\omega_0$.

\subsection{Frequency-Tuned WNPT Systems}
\label{sec:FreqTune}

Now let us consider frequency-tuned WNPT systems. In a frequency-tuned WNPT system, the source and load impedances are fixed. Generally, the coupled loops are conjugately matched to both source and load impedances at a fixed distance, and the system is frequency-tuned for different distances within this range.
The ratio of currents in two identical loops (see Fig. \ref{fig:BasicCkt}) can be written as:
\begin{equation}
i_2 = \frac{-j \omega M_{12}}{j\left(\omega L - \frac{1}{\omega C}\right) + \left( R + R_L \right)}i_1.
\label{eq:current_i2}
\end{equation}
According to (\ref{eq:SplitModes}), the even and odd mode frequencies are given by:
\begin{equation}
\left(\omega L - \frac{1}{\omega C}\right) = \pm \sqrt{\left(\omega M_{12}\right)^2 + \left(R+R_L\right)^2}.
\label{eq:evenodd2}
\end{equation}
The $+$ and $-$ solutions identify the modes above and below the resonant frequency ($\omega_0$) of the isolated loops, respectively. Substituting (\ref{eq:evenodd2}) into (\ref{eq:current_i2}), yields the following ratio of currents:
\begin{equation}
i_2 = \frac{i_1}{-j\frac{R+R_L}{\omega M_{12}} \pm \left[1 - \frac{\left(R + R_L\right)^2}{\left(\omega M_{12}\right)^2}\right]^\frac{1}{2}}.
\label{eq:current_i2i1}
\end{equation}
 It is clear from the expression above that the magnitude of the denominator of (\ref{eq:current_i2i1}) is equal to 1, and $\left|i_2\right|=\left|i_1\right|$.
Furthermore, an analysis of the phase of (\ref{eq:current_i2i1}) yields:
\begin{align}
\phi_+ &= \frac{R+R_L}{\omega M_{12}} \label{eq:phi1} \\
\phi_- &= -\pi - \frac{R+R_L}{\omega M_{12}}. \label{eq:phi2}
\end{align}

Therefore, the higher frequency mode ($\phi_+$) is the even mode, and the lower frequency mode ($\phi_-$) is the odd mode. It should be noted that these analytical findings were verified through circuit simulation in Agilent Advanced Design System (ADS).

Frequency-tuned systems maintain the frequency of operation at either the even or odd mode resonance in order to sustain a constant efficiency for distances within the strongly coupled region \cite{Intel, Korea, VehicalPower, FreqTune-Exp}. In these systems, the frequency is tuned to maintain (\ref{eq:SplitModes}), since $M_{12}$ changes with distance. The odd mode is generally preferred, since the currents in the loops are out of phase and their radiation cancels in the far field.

\subsubsection{Strongly Coupled WNPT System}

Operating at an even or odd mode frequency means that the system operates off of the resonant frequency, $\omega_0$, of the isolated loops. At an even or odd mode frequency, there exists an impedance mismatch and some power is reflected back to the source. It can be shown that the reflected power, proportional to (\ref{eq:ReflPower}), is minimized ($\frac{\delta \left| \Gamma_{\mathrm{in}} \right|^2}{\delta \omega} = 0$) for a fixed source/load impedance when operating at the even or odd mode resonance defined by (\ref{eq:SplitModes}). Therefore, operating at either the even or odd mode frequency guarantees the highest efficiency for a frequency-tuned WNPT system.

The reflection coefficient at the even or odd mode frequency can be found by substituting (\ref{eq:SplitModes}) into (\ref{eq:ReflPower}),
\begin{equation}
\left|\Gamma_{\mathrm{in}}\right|^2 =  \left(\frac{R}{R+R_L}\right)^2.
\label{eq:S_ReflPower}
\end{equation}
under the assumption that $\left(\omega M_{12}\right)^2 \ge \left(R+R_L\right)^2$. Substituting (\ref{eq:SplitModes}) and (\ref{eq:S_ReflPower}) into the expression for efficiency (\ref{eq:Efficiency2}) yields:
\begin{align}
\eta &=  \left(1-\left|\Gamma_{\mathrm{in}}\right|^2\right) \eta' \\
&= \left[1-\left(\frac{R}{R+R_L}\right)^2 \right] \left( \frac{R_L}{2R + R_L}\right) \\
&= \left(\frac{R_L}{R+R_L}\right)^2.
\label{eq:StrongCoupledEff}
\end{align}
Therefore, a constant efficiency is achieved by maintaining operation at either the even or odd mode resonant frequency of a strongly coupled WNPT system. The efficiency is not a function of distance (or equivalently  $M_{12}$), provided that (\ref{eq:SplitModes}) is upheld by properly tuning the frequency $\omega$ and ensuring that the loops are strongly coupled: $\left(\omega M_{12}\right)^2 \ge \left(R+R_L\right)^2$. Further, since $\Gamma_{\mathrm{in}}$ is constant, the input impedance is also constant for a frequency-tuned WNPT system.

\subsubsection{Critically and Weakly Coupled WNPT System}

Since critically and weakly coupled WNPT systems do not exhibit even or odd mode resonances, optimal efficiency is maintained by operating at $\omega_0$. The equations for the reflection coefficient  and total efficiency at critical and weak coupling are given by (\ref{eq:ReflPower_wkCrt}) and (\ref{eq:Efficiency_WkCrt}), respectively. Note that at critical coupling, $\left(\omega M_{12}\right)^2 = \left(R+R_L\right)^2$, and (\ref{eq:Efficiency_WkCrt}) reduces to (\ref{eq:StrongCoupledEff}). Thus, a smooth transition in efficiency occurs at critical coupling: the boundary between weak and strong coupling.

Finally, it should be noted that for fixed source and load impedances and operation at frequency $\omega_0$, the efficiency peaks at the distance (or equivalently the $M_{12}$ value) corresponding to critical coupling. In other words, $\frac{\delta \eta}{\delta M_{12}}=0$ when $\left(\omega M_{12}\right)^2=(R+R_L)^2 $ for  a fixed source and load impedance.

\subsection{Impedance-Tuned WNPT Systems}
\label{sec:CCImpMatch}

Next, let's explore a WNPT system that is simultaneously, conjugately matched to both source and load impedances for a specific distance of operation. Under these conditions, the optimal source and load impedances, $Z_{in}=Z_S^*$ and $Z_{out}=Z_L^*$, become \cite{Thomas}:
\begin{subequations}
\begin{align}
Z_S &= j \left(\frac{1}{\omega C_1}-\omega L_1\right) + \sqrt{{R_1}^2 + \frac{R_1}{R_2} \left(\omega M_{12}\right)^2}
\label{eq:OptZs} \\
Z_L &= j \left(\frac{1}{\omega C_2}-\omega L_2\right) + \sqrt{{R_2}^2 + \frac{R_2}{R_1} \left(\omega M_{12}\right)^2}.
\label{eq:OptZL}
\end{align}
\label{eq:OptZ}
\end{subequations}
At the resonant frequency $\omega_0$, these expressions reduce to:
\begin{subequations}
\begin{align}
R_S=\sqrt{{R_1}^2 + \frac{R_1}{R_2}\left(\omega_0 M_{12}\right)^2}
\label{eq:ZS_at_Res} \\
R_L=\sqrt{{R_2}^2 + \frac{R_2}{R_1}\left(\omega_0 M_{12}\right)^2}.
\label{eq:ZL_at_Res}
\end{align}
\label{eq:ZSZL_at_Res}
\end{subequations}
For identical loops, given by (\ref{eq: identicl}), the optimal source and load resistances at $\omega_0$ become:
\begin{equation}
R_{LS} = R_L=R_S = \sqrt{R^2 + \left(\omega_0 M_{12}\right)^2}.
\label{eq: Rlopt}
\end{equation}
Equation (\ref{eq: Rlopt}) defines an optimal value of $R_L=R_S$ at a given mutual inductance value (distance). If the system operates at $\omega_0$ and (\ref{eq: Rlopt}) is satisfied, the system exhibits the maximum possible efficiency:
\begin{subequations}
\begin{align}
\eta_{max} =  \eta' &= \left[ \frac{\omega_0 M_{12}}{R + R_{LS}} \right]^2 \\
&= \left[\frac{\omega_0 M_{12}}{R + \sqrt{R^2 + (\omega_0 M_{12})^2}} \right]^2.
\label{eq:Efficiency}
\end{align}
\end{subequations}
This expression for maximum efficiency is obtained by substituting (\ref{eq: Rlopt}) into either (\ref{eq:Efficiency_Base}) or (\ref{eq:Efficiency_WkCrt}). Equation (\ref{eq:Efficiency}) is the available gain ($G_A$) at the resonant frequency of the loops ($\omega_0$) \cite{Pozar}.

The expression for the optimal load resistance (\ref{eq: Rlopt}) can be rewritten as ${R_L}^2-R^2=\left(\omega M_{12}\right)^2$ and directly compared to the critical coupling condition: $\left(R+R_L\right)^2=\left(\omega M_{12}\right)^2$. Given that ${R_L}^2-R^2 < \left(R+R_L\right)^2$, a conjugately matched WNPT system is always weakly coupled. Since it is weakly coupled, the denominator of (\ref{eq:Efficiency}) is always less the numerator. In other words, the efficiency is less than unity.

\subsection{Discussion}
Subsection A showed that frequency tuning allows a constant efficiency to be maintained for distances within the strongly coupled range of operation. A higher source/load resistance results in a higher efficiency but also a shorter range of distances over which the system is strongly coupled. In other words, a higher efficiency can be maintained for close-in distances at the expense of a shorter range. Subsection B showed that maintaining a simultaneous conjugate match to source and load impedance through impedance tuning results in a weakly coupled system. Nonetheless, such an impedance matching scheme results in maximum possible efficiency. To summarize, a frequency-tuned WNPT system is simple to implement. An impedance-tuned system is more complex but can achieve higher efficiencies.

\begin{table}[htbp]
\centering
\begin{tabular}{|c|c|@{\hspace{-.005cm}}c@{\hspace{-.005cm}}|c|@{\hspace{-.005cm}}c@{\hspace{-.005cm}}|}
\hline
 \multicolumn{2}{|c|}{} & Strong & Critical & Weak \\
 \multicolumn{2}{|c|}{} & Coupling & Coupling & Coupling \\ \hline
 & \multirow{2}{*}{$|\Gamma|^2$} & \multirow{2}{*}{$\left(\frac{R}{R+R_L}\right)^2$} & Expressions & \multirow{2}{*}{$\left[ \frac{R^2-{R_L}^2 + \left(\omega M_{12}\right)^2 }
 {\left(R+R_L\right)^2 + \left(\omega M_{12}\right)^2} \right]^2$} \\
Frequency & & & are equal & \\ \cline{2-5}
Tuned & \multirow{2}{*}{$\eta$} & \multirow{2}{*}{$\left(\frac{R_L}{R+R_L}\right)^2$} & Expressions & \multirow{2}{*}{$\left[ \frac{2R_L \left(\omega M_{12}\right)}{\left(R+R_L\right)^2 + \left(\omega M_{12}\right)^2} \right]^2$} \\
 & & & are equal & \\ \hline
\multicolumn{2}{|c|}{Impedance-Tuned} & \multicolumn{3}{|c|}{$ \eta =\left[\frac{\omega_0 M_{12}}{R + \sqrt{R^2 + (\omega_0 M_{12})^2}} \right]^2  $} \\
\multicolumn{2}{|c|}{at $\omega_0$} & \multicolumn{3}{|c|}{} \\ \hline
\end{tabular}
\caption{Summary of the expressions for  $\left| \Gamma_{in} \right|^2$, and efficiency $\eta$, experienced under strong, critical, and weak coupling. The expressions for frequency-tuning are derived in section \ref{sec:FreqTune}, and those for impedance-tuning are derived in section \ref{sec:CCImpMatch}.}
\label{table:Freq_Imp_Comp}
\end{table}

\section{Analysis of Coupled Shielded Loop Resonators}
\label{sec:ShieldedLoop}

\begin{figure}[htbp]
    \centering
    \includegraphics[width=3.5in]{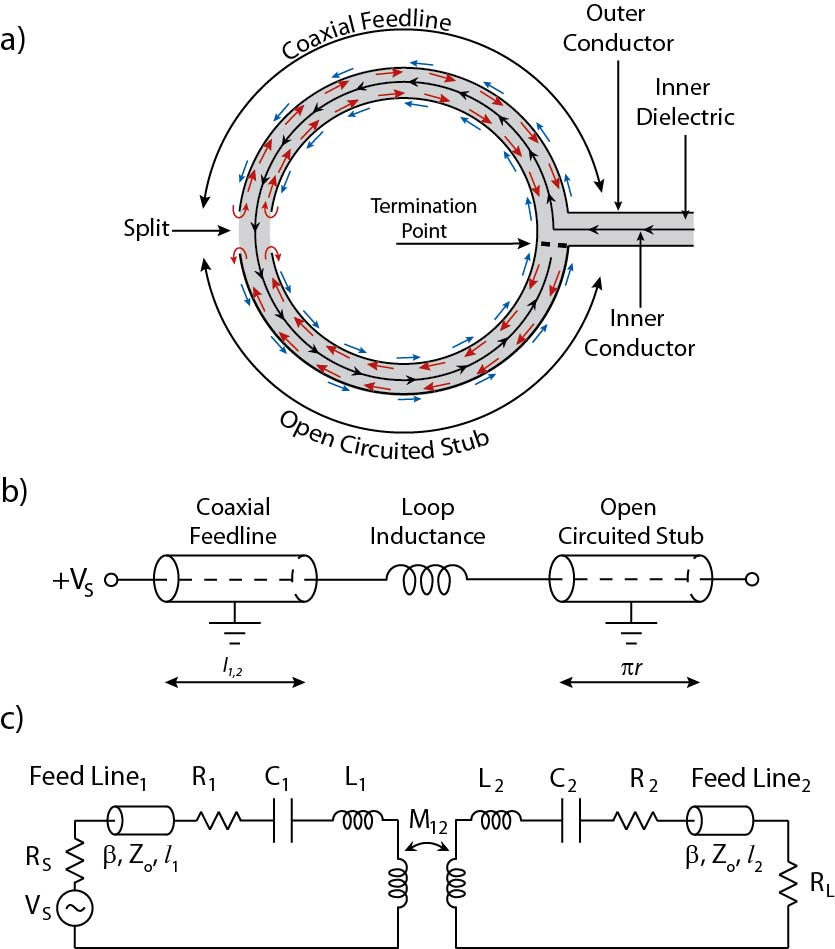}
    \caption{(a) Cross section of a shielded loop resonator. (b) Circuit model of an isolated, resonant shielded loop. (c) Circuit model for two magnetically coupled resonant shielded loops.}
    \label{fig:circuit}
\end{figure}

Resonant shielded loops will be used to experimentally investigate frequency-tuned and impedance-tuned WNPT. A shielded loop is a coaxial, electrically-small loop antenna with a primarily magnetic response \cite{ShieldedLoop-Stendgaard2, ShieldedLoop-Harpen}. The central conductor is left open circuited at the termination point of the loop to provide a resonance (see Fig. \ref{fig:circuit}a). A shielded loop  can be constructed from a semirigid coaxial cable by removing a small portion of the outer conductor at a point $\pi r$ from the feed point, where $r$ is the radius of the loop. This split in the outer conductor provides a current path which allows the open-circuited stub to be in series with the loop inductance. Currents supported by the resonant shielded loop are shown with arrows in Fig. \ref{fig:circuit}a.

Due to the skin depth, the currents on the outer surface of the loops are isolated from those within the coaxial cable comprising the loops \cite{ShieldedLoop-Stendgaard}. As a result, the resonant shielded loops can be broken down into a coaxial ``feed'' element, a loop inductance, and an open circuit transmission line. These three components are depicted in Fig. \ref{fig:circuit}b. Integrating these elements into the basic circuit model of Fig. \ref{fig:BasicCkt} yields the model shown in Fig. \ref{fig:circuit}c for magnetically coupled resonant shielded loops.

\begin{figure}[htbp]
    \centering
    \includegraphics[width=3.7in]{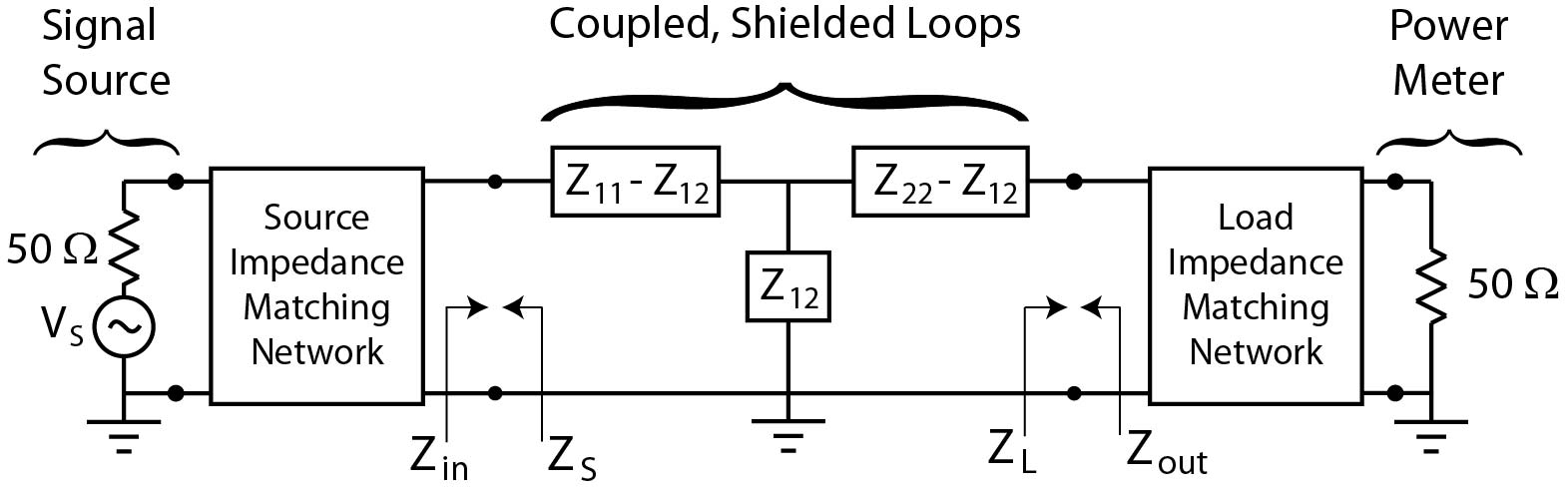}
    \caption{General T-network representing the magnetically coupled, resonant shielded loops. The elements are expressed in terms of the impedance matrix elements ($Z_{11},Z_{12},Z_{21},Z_{22}$) of the coupled loops. Source and load matching networks are shown connecting the loops to the 50 $\Omega$ signal source and power meter.}
    \label{fig:genTsect}
\end{figure}

A complete network representation for the circuit shown in Fig. \ref{fig:circuit}c can be derived by cascading the ABCD (transmission) parameters of the individual elements as follows:
\begin{gather}
    \begin{bmatrix}
        \cos \beta l_1 & jZ_o \sin \beta l_1 \\
        \frac{j}{Z_o} \sin \beta l_1 & \cos \beta l_1
    \end{bmatrix}
    \begin{bmatrix}
        1+\frac{Z_1}{Z_3} & Z_1+Z_2+\frac{Z_1Z_2}{Z_3} \\
        \frac{1}{Z_3} & 1+\frac{Z_2}{Z_3}
    \end{bmatrix}
    \notag \\
    *
    \begin{bmatrix}
        \cos \beta l_2 & jZ_o \sin \beta l_2 \\
        \frac{j}{Z_o} \sin \beta l_2 & \cos \beta l_2
    \end{bmatrix}.
    \label{eq:ABCDcascade}
\end{gather}
The first and last matrices in (\ref{eq:ABCDcascade}) represent the feedlines of the source and load loops, respectively. The central matrix represents the coupled loops shown in Fig. \ref{fig:circuit}c \cite{Pozar} with variables $Z_1$, $Z_2$, and $Z_3$ defined:
\begin{subequations}
\begin{align}
Z_1 &= Z_{11}-Z_{12} = R_1+\frac{1}{j\omega C_1}+j\omega \left(L_1-M_{12}\right) \label{eq:Z1} \\
Z_2 &= Z_{22}-Z_{12} = R_2+\frac{1}{j\omega C_2}+j\omega \left(L_2-M_{12}\right) \label{eq:Z2}\\
Z_3 &= Z_{12} = j\omega M_{12}. \label{eq:Z3}
\end{align}
\label{eq:Zexpressions}
\end{subequations}
The matrices in (\ref{eq:ABCDcascade}) are multiplied to obtain the complete transmission matrix for the system. This complete ABCD-matrix is then converted to an impedance matrix \cite{Pozar} and used to derive a T-equivalent circuit for the system, shown in Fig. \ref{fig:genTsect}. The optimal source and load impedances are then found by setting $Z_S={Z_{in}}^*$ and $Z_L={Z_{out}}^*$, respectively.

In \cite{Thomas}, two shielded-loop resonators were constructed and are used in this experimental study. The experimentally extracted parameters of the loops are given in Table \ref{table:LoopParameters}. The optimal $Z_S$ and $Z_L$ values were computed as a function of distance, or equivalently $M_{12}$ (see Fig. \ref{fig:smith}).
\begin{table}[htbp]
\centering
\begin{tabular}{|c|c|c|}
\hline
& Loop 1 & Loop 2  \\ \hline
L $(\mu H)$ & 0.596 & 0.583  \\ \hline
C $(pF)$ & 30.6 & 31.1 \\ \hline
R ($\Omega$) & 0.23 & 0.2  \\ \hline
Feedline Length (cm) & 38.5 & 39.5 \\ \hline
Radius (cm) & 10.7 & 10.7 \\ \hline
\end{tabular}
\caption{Circuit model parameters for two shielded-loop resonators \cite{Thomas}.}
\label{table:LoopParameters}
\end{table}

\begin{figure}[htbp]
    \centering
    \includegraphics[width=3.3in]{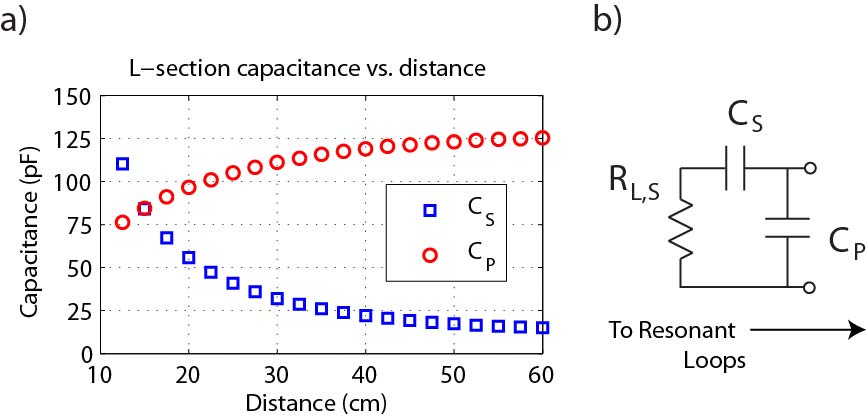}
    \caption{(a) Plot showing the capacitance values for the L-section matching networks needed to conjugately match the loops to 50 $\Omega$. (b) Schematic of the capacitive L-section matching networks.}
    \label{fig:capplot}
\end{figure}

Capacitive L-section matching networks were designed to match the $50$ $\Omega$ load and source to the coupled, shielded-loop system, as shown in Fig. \ref{fig:genTsect}. These fixed networks were designed for optimal power transfer at specific distances. The mutual inductance $M_{12}$ was computed as a function of distance for the two axially aligned electrically small loops using expressions for filamentary current loops given in \cite{Liepa}. The optimal capacitor values for the matching networks are plotted with respect to distance in Fig. \ref{fig:capplot}a. Optimal impedance values for $Z_S$ and $Z_L$ at distances of 20, 35, and 50 cm are shown in Table \ref{table:MatchValues}. Matching networks optimized for these three distances are referenced throughout this work.

\begin{table}[htbp]
\centering
\renewcommand{\arraystretch}{1.2}
\begin{tabular}{|c|c|c|}
\hline
Distance (cm) & $Z_S (\Omega)$ & $Z_L (\Omega)$  \\ \hline
20 & $5.94-30.15 j$ & $5.45-29.73 j$ \\ \hline
35 & $1.66-30.19 j$ & $1.52-29.76 j$ \\ \hline
50 & $0.76-30.2 j$ & $0.7-29.77 j$ \\ \hline
\end{tabular}
\caption{Calculated optimal impedance values at three discrete distances for the shielded loop WNPT system in \cite{Thomas}.}
\label{table:MatchValues}
\end{table}

\section{An Experimental Frequency-Tuned WNPT System Employing Resonant Shielded Loops}

In this section, the analysis from section \ref{sec:FreqTune} is validated experimentally. The shielded loop resonators and L-section matching networks from \cite{Thomas}, that conjugately match the coupled loops to $50$ $\Omega$ source and load impedances for distances of 20, 35, and 50 cm (values given in Fig. \ref{fig:capplot}a), were tested over a range of distances and frequencies. The loops were placed on an automated, 3-axis translation stage and the two-port scattering parameters were measured for the coupled, shielded loops (with impedance-matching networks in place) using a network analyzer (Hewlett Packard 3753D). The insertion loss associated with the fixed matching networks over the range of frequencies needed for frequency tuning ($\omega_0\pm\Delta\omega$) is negligible. Therefore, the efficiency is expressed directly as measured $|S_{21}|^2$.

\begin{figure}[htbp]
    \centering
    \includegraphics[width=3.75in]{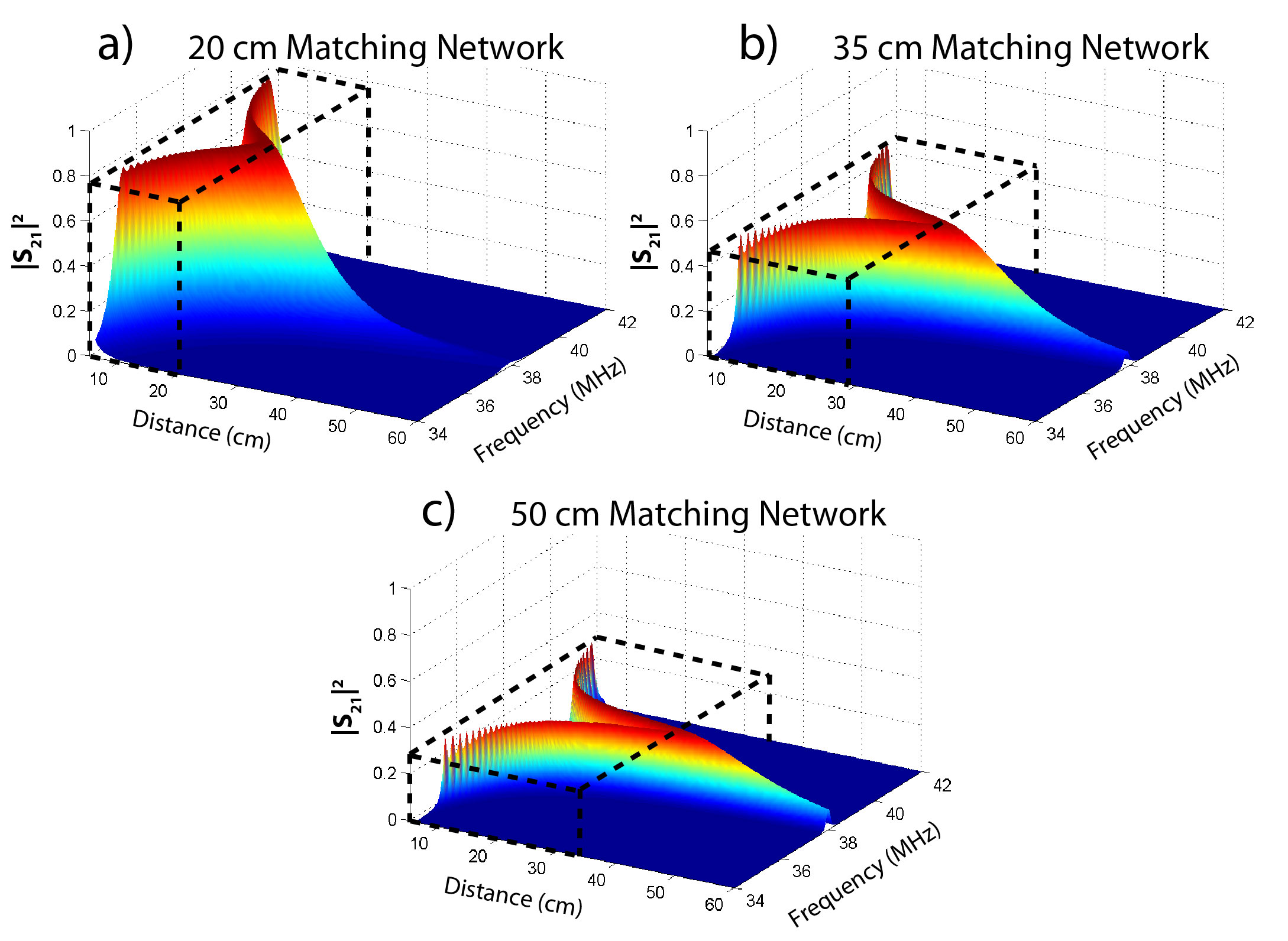}
    \caption{Experimental efficiency (${\left|S_{21}\right|}^2$) vs. distance and frequency for two magnetically coupled, resonant shielded loops. The coupled loops were impedance-matched to the 50 $\Omega$ source and load using matching networks designed for (a) 20 cm, (b) 35 cm, and (c) 50 cm loop separations. The separation into even and mode frequencies is evident as the two loops become strongly coupled. Dotted lines denote the regions of strong coupling.}
    \label{fig:S21}
\end{figure}

For each pair of matching networks, Fig. \ref{fig:S21} displays the experimental efficiency ($|S_{21}|^2$) as a function of frequency and distance. The volume encompassed by the dotted lines denotes the strongly coupled region corresponding to each matching network. Note that similar characteristics were reported in \cite{Intel} for a single load. By testing several different matching networks, one can truly see the effect that the static-load has on the overall efficiency of the system.

\begin{figure}[htbp]
    \centering
    \includegraphics[width=3.6in]{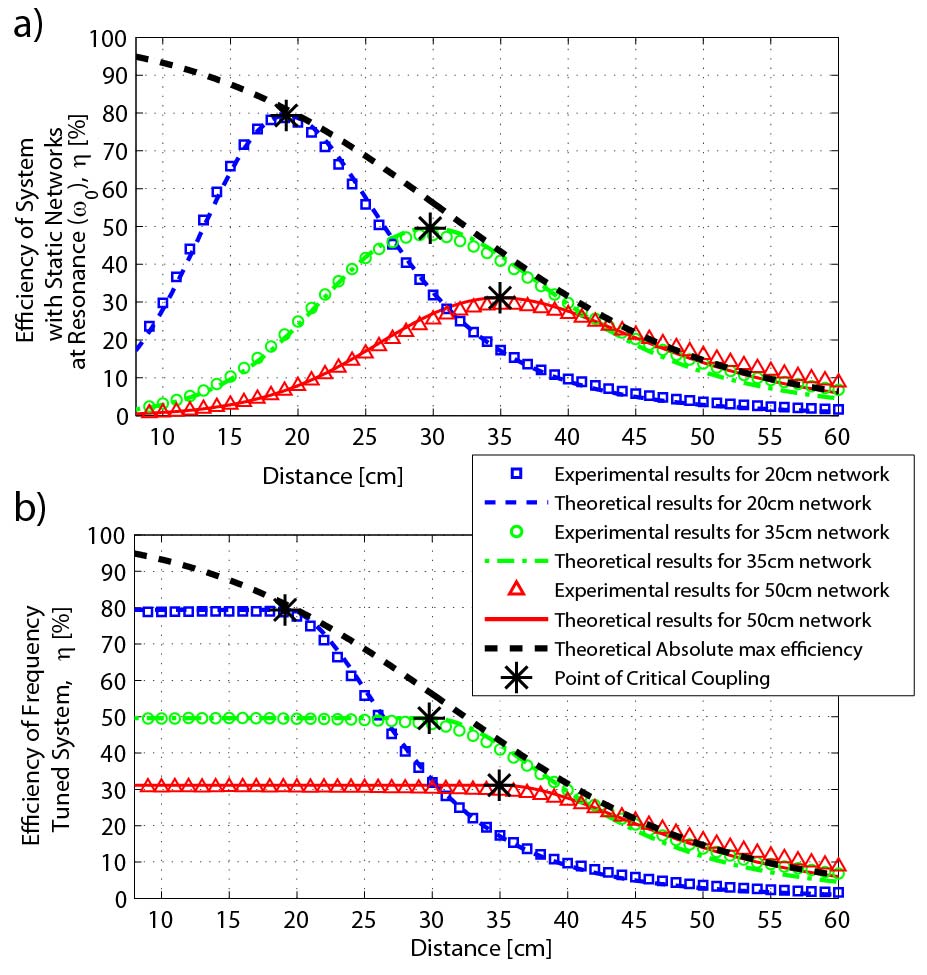}
    \caption{Plots of experimental and theoretical efficiencies vs. distance using fixed L-section matching networks. (a) The experimental and theoretical efficiencies at the self-resonant frequency of the loops ($\omega_0$). (b) The experimental and theoretical efficiencies under frequency-tuning. While in strong coupling, the system was frequency-tuned to operate at an even or odd mode frequency in order to maintain a constant efficiency, as in (\ref{eq:StrongCoupledEff}). The $\square, \bigcirc$, and $\triangle$ plot the experimental efficiencies (${\left|S_{21}\right|}^2$) vs. distance for three fixed matching networks designed for loop separations of 20, 35 and 50 cm, respectively. The dotted, dash-dotted, and solid lines plot the corresponding theoretical efficiencies. In (a), theoretical efficiencies were calculated using (\ref{eq:Efficiency_WkCrt}). In (b), efficiencies were calculated for weak coupling using (\ref{eq:Efficiency_WkCrt}), and for strong coupling using (\ref{eq:StrongCoupledEff}). The dashed line plots the maximum achievable efficiency given by (\ref{eq:Efficiency}). Distances corresponding to critical coupling are marked with an asterisk.}
    \label{fig:S21_v_Pow}
\end{figure}

The experimental data shown in Fig. \ref{fig:S21} was used to plot the efficiency curves of Fig. \ref{fig:S21_v_Pow}. Fig. \ref{fig:S21_v_Pow}a shows the experimental efficiency at the resonant frequency $\omega_0$ versus distance, and compares it to theory. The theoretical curves were generated using (\ref{eq:Efficiency_WkCrt}), the loop parameters given in Table II, and the source and load impedances for the three distances given by (\ref{eq:ZSZL_at_Res}). A loss of efficiency due to reflections occurs for distances less than the critical coupling point. Fig. \ref{fig:S21_v_Pow}b shows the experimental efficiency when frequency tuning is employed, and compares it to theory. The theoretical curves were generated using (\ref{eq:StrongCoupledEff}) for distances of strong coupling, and (\ref{eq:Efficiency_WkCrt}) for distances of critical and weak coupling. For critical and weak coupling distances, the system operates at the resonant frequency $\omega_0$. Once again, the actual loop parameters given in Table \ref{table:LoopParameters}, as well as the source and load impedances given by (\ref{eq:ZSZL_at_Res}) for the three distances, were used in the theoretical calculations. Table \ref{table:ExpTheory_Comp} compares the experimental efficiencies for strong coupling to the theoretical values given by (\ref{eq:StrongCoupledEff}).

For comparison purposes, Figs. \ref{fig:S21_v_Pow}a and \ref{fig:S21_v_Pow}b also show the maximum possible efficiency vs. distance given by (\ref{eq:Efficiency}), which assumes a simultaneous conjugate match at all distances. The expected critical coupling distances are also labeled in both figures. These points were determined by finding the distance where $(\omega M_{12})^2 = (R + R_L)^2$ for each of the three matching networks.

\begin{table}[htbp]
\centering
\begin{tabular}{|c|c|c|c|}
\hline
                & {Experimental}                    & {Theoretical}                         &  {Predicted} \\
 {Matching}     & {Results}                         & {Results}                             &  {Critical} \\
 {Network}      & {from Fig. \ref{fig:S21_v_Pow}}   & {from (\ref{eq:StrongCoupledEff})}   & {Coupling Point} \\ \hline
20 cm (1.87$r$) & $78.85\%$                         & $79.37\%$                             & 19.13 cm \\ \hline
35 cm (3.27$r$) & $49.5\%$                         & $49.58\%$                              & 29.79 cm \\ \hline
50 cm (4.67$r$) & $30.54\%$                         & $31.14\%$                              & 34.95 cm \\ \hline
\end{tabular}
\caption{A comparison of experimental, frequency-tuned efficiencies and those theoretically found for strong coupling using (\ref{eq:StrongCoupledEff}).}
\label{table:ExpTheory_Comp}
\end{table}

Therefore, by tuning to the even or odd mode frequency of the strongly coupled shielded loop system, constant experimental efficiencies are observed. These experimental efficiencies are in excellent agreement with theory. Nevertheless, the efficiencies attained through frequency-tuning are lower than those that can be attained through a complex conjugate match (labeled Theoretical Absolute max efficiency in  Fig. \ref{fig:S21_v_Pow}).

It should be noted that the further the distance at which the system is matched, the lower the optimal $R_L$ will be. Selection of a lower $R_L$ results in lower efficiencies at close distances but extends the distances over which the system is strongly coupled (see Table \ref{table:ExpTheory_Comp}). This expands the range over which frequency-tuning applies, and results in higher efficiencies at further distances.

\section{An Experimental Impedance-Tuned WNPT System Employing Varactor-Based Matching Networks}
\label{sec:LP_Varactors}

A complex conjugate match, given by (\ref{eq: Rlopt}), can be satisfied for a given $M_{12}$ using fixed matching networks. However, as $M_{12}$ changes (relative position of the loops varies), this condition is no longer maintained. In this section, the analysis of the impedance-tuned system presented in section \ref{sec:CCImpMatch} is validated experimentally through the use of tunable impedance-matching networks.

\begin{figure}[htbp]
    \centering
    \includegraphics[width=1.5in]{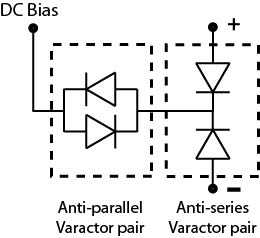}
    \caption{Linearized topology suggested in \cite{Buis1, Buis2}. The anti-series varactor pair functions as a single variable capacitor, and an anti-parallel varactor pair provides a high zero-bias impedance for the DC biasing supply as in \cite{Buis3}.}
    \label{fig:APASnet}
\end{figure}

The efficacy of varactor-based impedance-matching networks has been demonstrated in earlier works \cite{VaracMatch, VaracAmps}. Here, varactors are used to implement tunable L-section matching networks for a WNPT system. The capacitance of a varactor varies directly with an applied reverse voltage bias. When a RF signal is applied atop this bias, non-linearities due to harmonic distortion can occur. Earlier works have analyzed this \cite{Meyer-Varacts,Buis1,Buis2}, and proposed linearized topologies to mitigate these effects. In these proposed solutions, an anti-parallel varactor pair forms the feed point for DC-bias and ground, and an anti-series varactor pair forms the variable capacitor. These networks are desired for third harmonic suppression in order to reduce distortion. For the suppression to occur, a diode grading coefficient of $M\approx0.5$ is desired \cite{Meyer-Varacts}. A low-distortion varactor network in the form of this anti-parallel anti-series combination was used in this work (Fig. \ref{fig:APASnet}).

\subsection{Circuit Development}

This section describes how varactor-based tunable capacitors were integrated into L-section matching networks to yield a tunable WNPT system. The total capacitance of a diode is given by:
\begin{equation}
    C_V = \frac{C_{jo}}{{\left(1+\frac{V_R}{V_J}\right)}^M}+C_P,
    \label{eq:CV}
\end{equation}
where $C_{jo}$ is the junction capacitance of the diode, $V_R$ is the applied reverse voltage (the biasing voltage), $V_J$ is the built-in diode junction voltage, $C_P$ is the package capacitance, and $M$ is the grading coefficient.

\begin{figure}[htbp]
    \centering
    \includegraphics[width=3.5in]{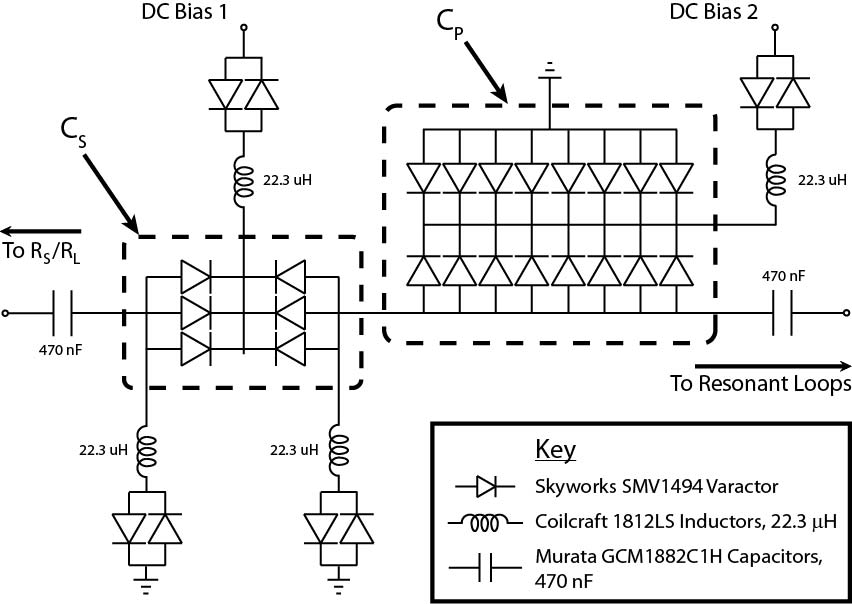}
    \caption{A circuit schematic of the varactor-based L-section matching network. Coilcraft 1812LS 22.3 $\mu$H inductors were used as RF chokes to isolate the RF signal from the bias voltage and ground. Murata GCM1882C1H 470 nF capacitors serve as DC blocks, to constrain the biasing voltage from the resonant loops. The anti-series varactor pairs acting as matching capacitors $C_S$ and $C_P$ (see Fig. \ref{fig:capplot}b) are noted on the image. All varactors are Skyworks SMV1494 abrupt junction varactors.}
    \label{fig:CktSchem}
\end{figure}

By placing several anti-series varactor pairs in parallel, the equivalent capacitances needed to meet the matching network criteria were achieved (see Fig. \ref{fig:capplot}a). An abrupt-junction varactor diode (Skyworks SMV1494) was selected. This varactor has a grading coefficient $M=0.47$, thus meeting the criteria for linear operation. In addition, the SMV1494 has $C_{jo}=58$ pF and comes in the SC-79 package for reduced printed circuit board footprint, low parasitic inductance, and a nearly negligible package capacitance ($C_P\approx0$ pF). The number of anti-series varactor pairs placed in parallel was chosen to be 3 and 8 for $C_S$ and $C_P$, respectively (see Fig. \ref{fig:CktSchem}). This situates the reverse bias voltage ($V_R$) required to properly bias the networks within the operational range of the SMV1949 varactors ($0\sim15 $ V). The 3 anti-series varactor pairs comprising $C_S$ were able to achieve a tunable capacitance ranging from 86.7 pF to  to 22 pF, for $V_R=0$ V to $V_R=10$ V, respectively. The 8 anti-series varactor pairs comprising $C_P$ had a tunable range of 232 pF to 58.8 pF, for $V_R=0$ V to  at $V_R=10$ V, respectively. The final circuit schematic of the matching network is shown in Fig. \ref{fig:CktSchem}.

\subsection{Experimental Validation}
\label{sec:ExpResults}

\begin{figure}[htbp]
    \centering
    \includegraphics[width=3.5in]{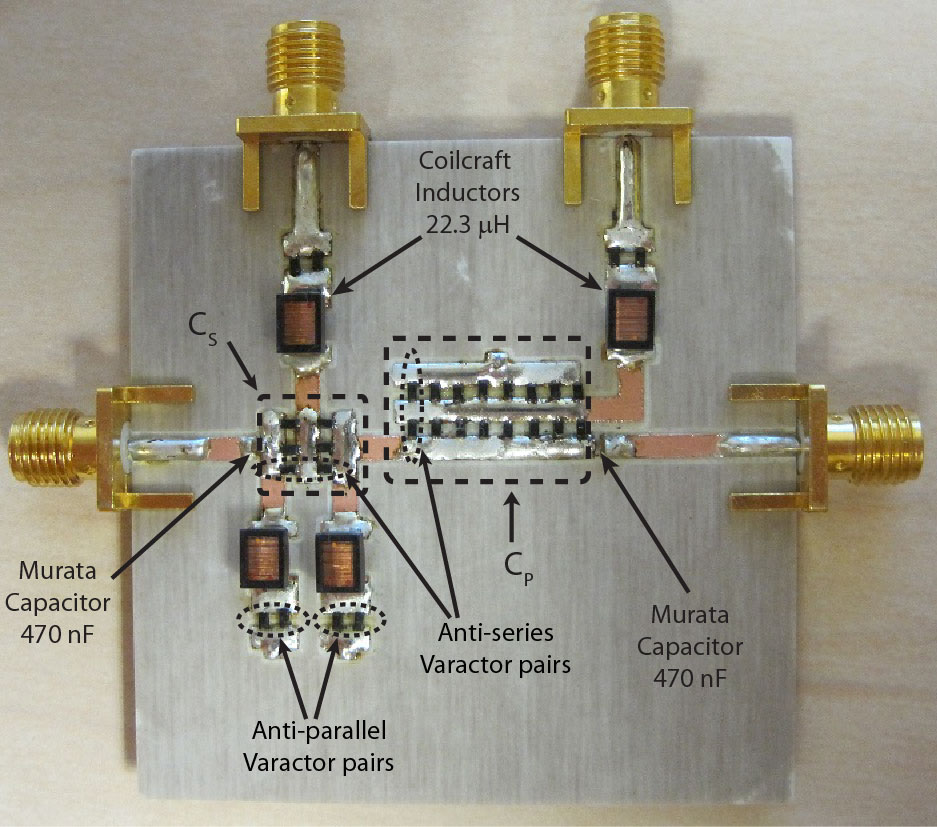}
    \caption{Photograph of a varactor-based L-section matching network. This is the fabricated realization of the circuit schematic shown in Fig. \ref{fig:CktSchem}. Further component data is given in \cite{Components}.}
    \label{fig:CktImage}
\end{figure}

The circuit in Fig. \ref{fig:CktSchem} was simulated using the Agilent Advanced Design System (ADS) harmonic balance solver. One of the two fabricated varactor-based L-section matching networks is depicted in Fig. \ref{fig:CktImage}.
Two such tunable matching networks were tested on the shielded-loop system. An analog signal generator (Agilent N5183A) was used to supply 10 mW of power at 38 MHz to the source loop. The output power was measured with a power meter (Agilent E4416A) and power sensor (Agilent N8485A), as depicted in Fig. \ref{fig:genTsect}. The two DC bias points on the networks were tuned independently with a DC voltage supply (Agilent E3648A/E3631A).

\begin{figure}[htbp]
    \centering
    \includegraphics[width=3.65in]{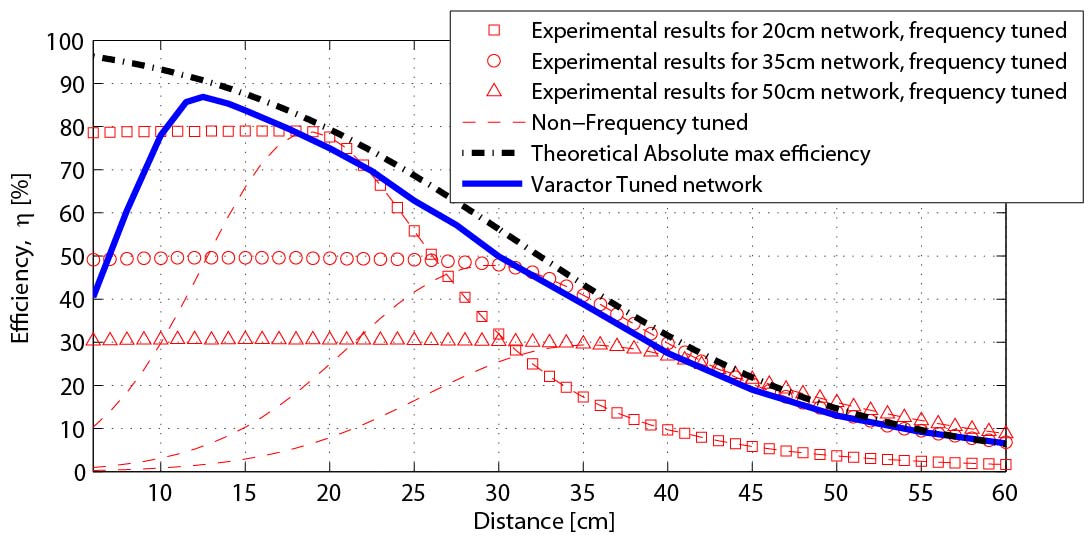}
    \caption{A plot of efficiency vs. distance for the fixed-capacitor and varactor-based tunable L-section matching networks. The solid line indicates the results from the varactor network. The $\square$, $\bigcirc$, and $\triangle$ plots show the results from Fig. \ref{fig:S21_v_Pow} for comparison. The dash-dot line plots the theoretical maximum achievable efficiency given by (\ref{eq:Efficiency}).}
    \label{fig:LPResults}
\end{figure}

The theoretical DC bias values were derived by interpolating the required capacitances vs. distance from Fig. \ref{fig:capplot}a with the $V_R$ vs. $C_V$ characteristics of the varactor given by (\ref{eq:CV}). The experimental efficiency of the impedance-tuned, coupled shielded loops is plotted with respect to distance in Fig. \ref{fig:LPResults}. It is compared to the efficiency curves for the fixed-capacitor matching networks (see Fig. \ref{fig:S21_v_Pow}).
The experimental efficiency is approximately $3\%$ lower than the theoretical maximum efficiency assuming a perfect conjugately matched system given by (\ref{eq: Rlopt}). This slight loss in efficiency is due to parasitics, tolerances, and losses of the manually constructed varactor-based matching networks.

\begin{figure}[htbp]
    \centering
    \includegraphics[width=2.3in]{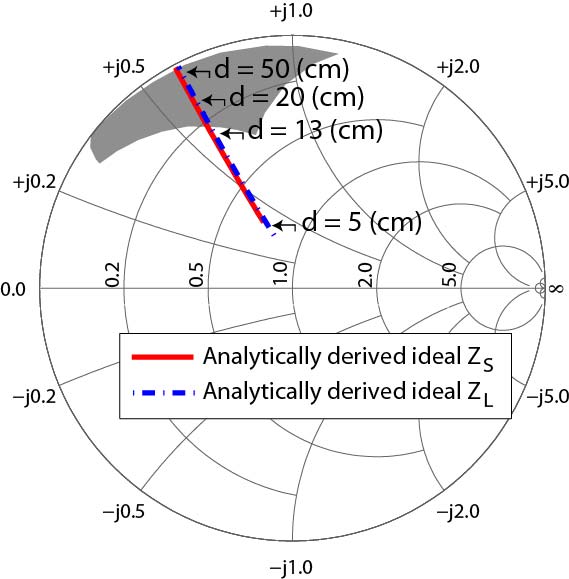}
    \caption{Plotted on the Smith chart are the optimal $Z_S$ and $Z_L$ values for varying distances, given by (\ref{eq:OptZs}) and (\ref{eq:OptZL}) respectively. For the shielded loops presented in \cite{Thomas}, the values are nearly the same, indicating identical loops. The shaded grey region represents the impedances that can be matched by the varactor network used in this work.}
    \label{fig:smith}
\end{figure}

Fig. \ref{fig:smith} plots the optimal load and source impedance values for a conjugately matched shielded-loop system (as they were derived in section \ref{sec:ShieldedLoop}). The shaded grey area is the region that can be matched by the varactor-based L-section matching networks used in this work. The required source/load impedances stray from this region for distances $<13$ cm, giving rise to an impedance mismatch. This explains the significant drop in efficiency at these distances (see Fig. \ref{fig:LPResults}). This could be mitigated by using varactors with a larger tuning range, or using a modified network that is tailored to a different range of distances. Alternatively, for distances $<13$ cm, the tunable network can be biased to establish a strongly coupled system (satisfying $\omega M_{12} > R+R_L$) and frequency-tuning employed to maintain a constant efficiency at close-in distances.

In this section, it was shown that by maintaining a conjugate impedance-match with tunable varactor-based matching networks, one can achieve higher efficiencies than with a frequency-tuned WNPT system. This however comes at the cost of added complexity.

\subsection{Angular and Axial Misalignment}

\begin{figure}[htbp]
    \centering
    \includegraphics[width=2.5in]{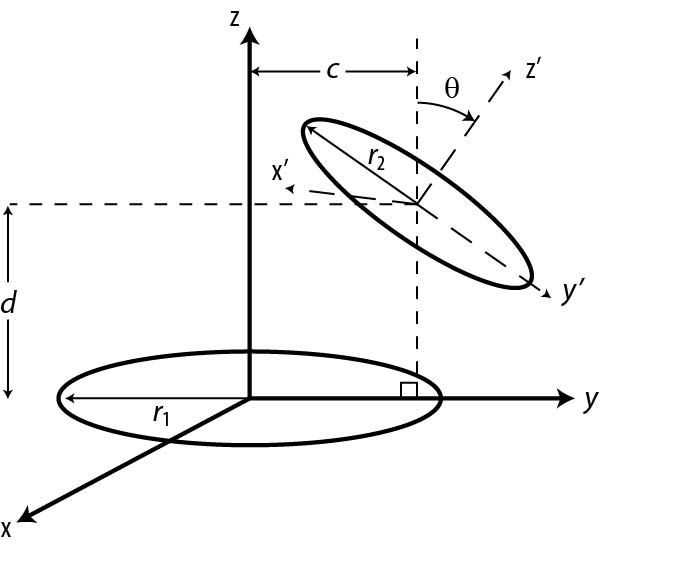}
    \caption{Relative position of the resonant shielded loops with respect to each other. The plots identifies variations in axial distance, axial misalignment and angle between the two planes of the loops.}
    \label{fig:MutualLoops_All}
\end{figure}

Up to this point, mutual inductance $(M_{12})$ was varied only as a function of axial distance. In reality, $M_{12}$ is affected by variations in axial distance ($d$), axial misalignments ($c$), and the angle ($\theta$) between the two planes of the loops, as shown in Fig. \ref{fig:MutualLoops_All}. These variations all affect the efficiency of power transfer.

Using \cite{Grover, Mutual}, analytical values for $M_{12}$ under these variations were found, and the corresponding efficiencies were computed. Theoretical and experimental results were obtained by aligning the loops co-axially at one of three discrete distances ($d=$ 20, 35, or 50 cm). They were then further subjected to either:
\begin{itemize}
\item Axial misalignments ($d$ fixed, $\theta=0$, and $c$ varied)
\item Angular variations ($d$ fixed, $c=0$, and $\theta$ varied)
\end{itemize}
\begin{figure}[htbp]
    \centering
    \includegraphics[width=3.2in]{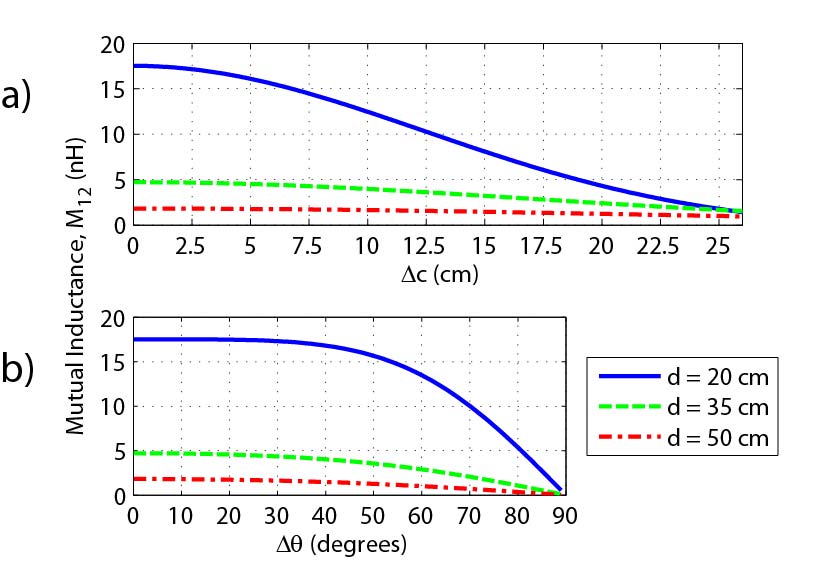}
    \caption{Plots show $M_{12}$ at 38 MHz as a function of (a) axial misalignment ($\Delta c$) of parallel loops (b) angular misalignment ($\Delta \theta$) between the two loops for 3 different axial distances ($d$).}
    \label{fig:M12_AngleAxial}
\end{figure}

The plots in Figs. \ref{fig:M12_AngleAxial}a and \ref{fig:M12_AngleAxial}b depict the changes in $M_{12}$ for movements off-axis ($\Delta c$) and changes in angle ($\Delta \theta$), respectively. Note that at far distances ($d \gg r$), $\Delta M_{12}$ is much smaller for changes in $c$ and $\theta$ than at close distances.
\begin{figure}[htbp]
    \centering
    \includegraphics[width=3.6in]{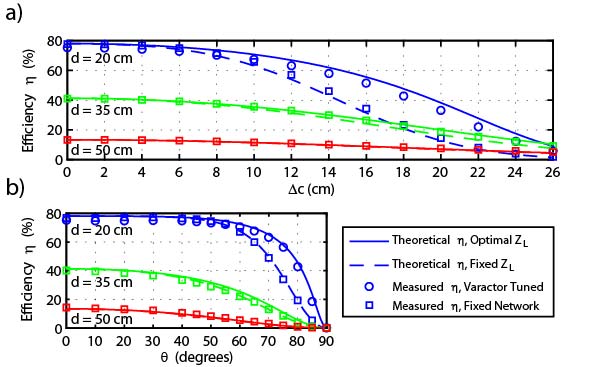}
    \caption{Power transfer efficiency as a function of (a) distance off-axis ($\Delta c$) for parallel loops, and (b) rotation ($\Delta \theta$) between two loops aligned along a central axis. The solid lines depict theoretical maximum efficiencies with the assumption that optimal source (\ref{eq:OptZs}) and load (\ref{eq:OptZL}) are used. The dotted lines depict the theoretical efficiencies for resonant shielded-loop WNPT system employing fixed-capacitor matching networks optimized for 20, 35, or 50 cm distances. The  $\square$ plots represents measured efficiencies as the system with fixed-capacitor matching networks undergoes axial or angular misalignments. The $\bigcirc$ plots represents measured efficiencies as the system with varactor-based matching networks undergoes axial or angular misalignments and the networks are tuned for maximum efficiency. Distances ($d$) are fixed for each set of measurements, and are denoted on the plots.}
    \label{fig:AngleAxialResults}
\end{figure}

The experimental results in Fig. \ref{fig:AngleAxialResults} depict the changes in efficiency ($\eta$) under the conditions described above. Results for both fixed-capacitor matching networks and tunable varactor-based networks are in good agreement with theory. At $d=$ 35 and 50 cm, the efficiencies using fixed-capacitor networks are nearly the same as the maximum efficiency, since changes in $M_{12}$ are negligible. Therefore, results for the tunable network are only displayed at 20 cm, where they can adapt to changes in mutual inductance, $M_{12}$, and show appreciable gains in efficiency.

\section{A Higher Power Matching Network}

In the experiments of section \ref{sec:LP_Varactors}, 10 mW of power was supplied to the tunable matching networks and resonant shielded loops. In practice, it is necessary to scale power levels to those used by modern electronic devices. In this section, tunable varactor-based matching networks capable of handling higher power levels are considered. Varactor networks with high power handling capability have been demonstrated in \cite{VaracAmps,Buis1}.

Commercially available Micrometrics MTV4045-10 and MTV4060-16 abrupt junction tuning varactors were selected for capacitances $C_S$ and $C_P$ (see Fig. \ref{fig:capplot}a), respectively. These varactors have high breakdown voltages $(B_V)$ of 45 V and 60 V, respectively, and a grading coefficient of $M=0.46$.

\begin{figure}[htbp]
    \centering
    \includegraphics[width=3.4in]{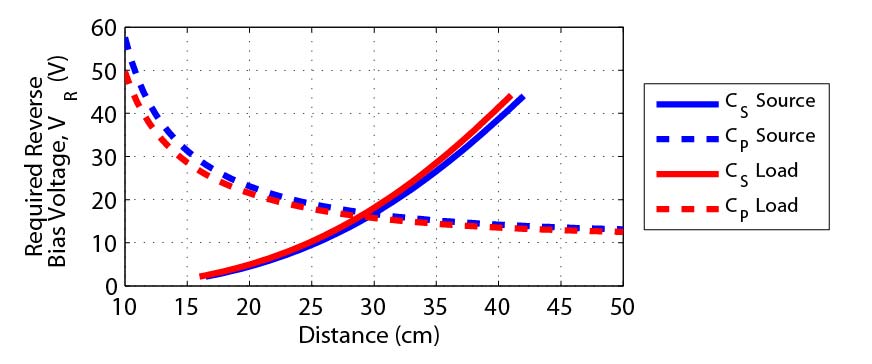}
    \caption{The capacitances required as a function of distance for the varactor-based L-section matching networks. The $C_V$ vs. The curves show the theoretical reverse bias voltages needed to achieve the required capacitances for the high power matching networks.}
    \label{fig:InterpResultsHP}
\end{figure}

It was determined through simulation and interpolation that 44 and 40 pairs of anti-series varactor pairs met the requirements from for $C_S$ and $C_P$, respectively (see Fig. \ref{fig:capplot}a). The results of the interpolation can be seen in Fig. \ref{fig:InterpResultsHP}. Simulations showed that the network could handle an applied power of approximately $5$ W. Beyond this power level, the varactor specifications are exceeded and harmonic distortion degrades the WNPT system's performance.

\begin{figure}[htbp]
    \centering
    \includegraphics[width=3.6in]{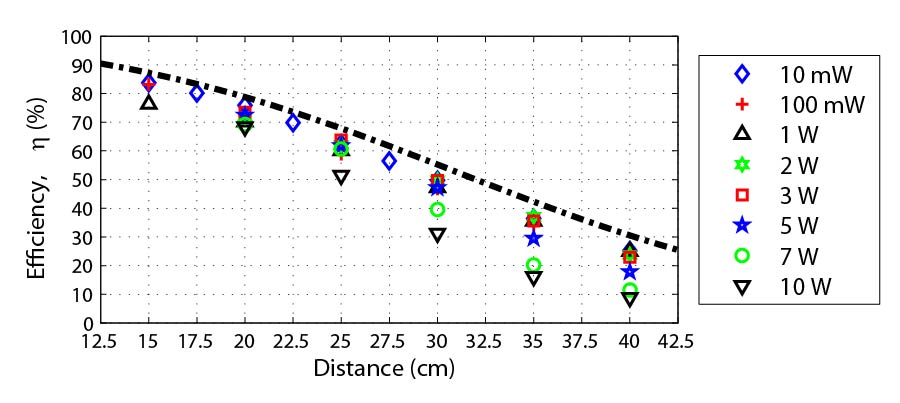}
    \caption{A plot of efficiency vs. distance for the resonant shielded-loop WNPT system employing high-power varactor-based L-section matching networks for various source power levels. The dash-dot line plots the theoretical maximum achievable efficiency given by \ref{eq:Efficiency}.}
    \label{fig:HPResults}
\end{figure}

These higher power L-section matching networks were tested on the shielded loops. The 38 MHz signal from the source (Hewlett Packard 8654) was amplified by a Minicircuits 52 dB RF Class A amplifier (ZHL-ED12128A/1) to produce the desired input power level $P_{in}$. The output power was measured with a power meter (Agilent E4416A) and high-power sensor (Agilent N8481). The varactor-based L-section matching networks were independently tuned with a DC voltage source (Agilent E3648A/E3631A). The experimental efficiencies of the high power network over distance are shown in Fig. \ref{fig:HPResults}, for multiple input power levels.

The results demonstrate the ability of the varactor-based L-section matching networks to provide an impedance-match for varying values of $M_{12}$, in order to maximize efficiency. However, once the input power is increased beyond 5 W, a decrease in the experimental efficiencies occurs. This is due to a few reasons. First, the varactor specifications are exceeded; most notably the maximum current ratings. In addition, harmonic distortions begin to affect efficiency at these higher power levels. Furthermore, an impedance mismatch is incurred at farther distances. Fig. \ref{fig:InterpResultsHP}, shows that the required reverse bias voltage $V_{R}$ approaches $B_V=45$ V (breakdown voltage for the MTV4045-10) as the distance is increased. Therefore, $V_R$ was lowered as $V_{RF}$ increased to ensure that $V_R + V_{RF} < B_V$ \cite{VaracAmps}. This in turn resulted in a higher than desired capacitance $C_S$. Using a varactor with a higher $B_V$ would alleviate this \cite{Components2}. Varactors with high linearity and tuning range \cite{SiG1}, as well as varactors with higher power handling capabilities \cite{SiC} have been proposed in the past, but are not commercially available.

\section{Conclusion}

In this paper, frequency tuning and impedance tuning techniques were explored for increasing the efficiency of wireless non-radiative power transfer (WNPT) systems. Both techniques were investigated analytically using a lumped-element circuit model. Expressions for the input reflection coefficients and efficiencies were derived. It was shown that a frequency-tuned system can maintain constant efficiency in the strongly coupled region of operation. This constant efficiency, however, is limited by reflections which prevent optimal operation. In addition, an impedance-tuned WNPT system was investigated. It was shown that impedance-matched WNPT system can achieve optimal/maximum possible efficiency.

Both frequency tuning and impedance tuning were experimentally demonstrated using a resonant shielded-loop WNPT system. The shielded-loop system exhibited even and odd modes in the strong coupling region. By tuning the operating frequency to one of these modes, a constant efficiency was achieved for distances within the strongly coupled region.  The experimental impedance-tuned WNPT system employed varactor-based matching networks to achieve maximum efficiency. The matching networks maintained a simultaneous conjugate match to source and load over a range of distances. Further, it was shown that such matching networks can compensate for changes in the mutual inductance, $M_{12}$, which occur with axial misalignments and angular variations of the transmitter and receiver loops.

Finally, higher power matching networks were developed and experimentally demonstrated for the impedance-tuned WNPT system. The matching networks were realized using multiple varactors to withstand power levels exceeding 5 W. A reduction in efficiency due to device limitations and harmonic distortion was observed as power levels increased. A varactor with a higher reverse bias voltage could alleviate these issues.

Control circuity can be developed to create a smart WNPT system capable of dynamically improving efficiency over a wide range of practical distances. This smart system could employ dynamic frequency tuning \cite{Auto-Tune}, impedance tuning or a combination of both. For example, a switch matrix could switch in or out static matching networks designed for various fixed distances. Frequency tuning could then be employed for distances between these fixed values.
\bibliographystyle{IEEE}
\bibliography{Biblio}

\begin{thebibliography}{10}
\bibitem{Components2}
``{Note on Components},'' {MTV4045-10 and MTV4060-16 varactors have a reverse
  breakdown voltage $B_V=$ 45 V and 60 V, respectively. The MTV4090 series has
  a breakdown voltage of $V_B=$ 90 V, and is desired. However, the varactors in
  this study were chosen for their product-line availability}.
\end{thebibliography}
\end{document}